\documentclass[fleqn,usenatbib]{mnras}

\usepackage{newtxtext,newtxmath}
\usepackage{color}

\usepackage[T1]{fontenc}

\DeclareRobustCommand{\VAN}[3]{#2}
\let\VANthebibliography\thebibliography
\def\thebibliography{\DeclareRobustCommand{\VAN}[3]{##3}\VANthebibliography}

\usepackage{graphicx}	
\usepackage{amsmath}	
\usepackage{multirow}
\usepackage{booktabs}

\title[EN1: UVIT catalogue and star formation]{Deep far-UV observations of the ELAIS N1 field using AstroSat: Source catalogue, spectral energy distribution modelling and star formation}

\author[]{
Pranjal Chaturvedi,$^{1}$\thanks{E-mail: phd2301121012@iiti.ac.in, pranjalchaturvedi@outlook.com}
Akriti Sinha,$^{2}$
Abhirup Datta,$^{1}$
and Kanak Saha$^{3}$
\\
$^{1}$ Department of Astronomy, Astrophysics and Space Engineering, Indian Institute of Technology Indore, Indore, 453552, India \\
$^{2}$ Max Planck Institute for Solar System Research, Justus-von-Liebig-Weg 3, 37077 Göttingen, Germany\\
$^{3}$ Inter-University Centre for Astronomy and Astrophysics, Ganeshkhind, Post Bag 4, Pune 411007, India
}

\date{Accepted XXX. Received YYY; in original form ZZZ}

\pubyear{2026}

\begin{document}
\label{firstpage}
\pagerange{\pageref{firstpage}--\pageref{lastpage}}
\maketitle

\begin{abstract}
We present a far-ultraviolet (FUV) photometric study of the ELAIS N1 deep field using the Ultra-Violet Imaging Telescope (UVIT) onboard AstroSat, observed in the F154W filter ($\lambda_{\rm eff} = 1541$\,\AA) with a total on-source exposure time of 30\,ksec. Level 1 data were reduced using CCDLAB v3.0, yielding source catalogues of 1637 objects at $3\sigma$ and 458 objects at $5\sigma$, with limiting magnitudes of $25.69\,m_{AB}$ and $25.13\,m_{AB}$ respectively. FUV positions are cross-matched against multiwavelength catalogues spanning optical and infrared wavelengths, with redshifts drawn from spectroscopic and photometric sources. Active galactic nuclei (AGN) are identified and excluded via established multiwavelength criteria, leaving a clean sample of star-forming galaxies (SFGs). Spectral energy distribution (SED) modelling is performed using CIGALE, employing a delayed star formation history with an optional late burst, Bruzual \& Charlot stellar population synthesis, Calzetti dust attenuation, and the SKIRTOR AGN module. From the best-fit models, we derive star formation rates (SFRs), total stellar masses, and young stellar masses as a function of redshift. The SFR increases monotonically with redshift, consistent with the evolution of the Star Formation Main Sequence (SFMS). The ratio of young-to-total stellar mass remains approximately constant across $0 < z \lesssim 0.76$, confirming that the sample consists predominantly of secularly evolving systems undergoing steady, self-regulated star formation rather than starburst-driven episodes.

\end{abstract}




\section{Introduction}

The ELAIS-N1 (European Large-Area ISO Survey-North 1) field is a premier extragalactic deep field, centered at approximately $RA = 16^h 10^m, Dec = +54^\circ 30'$, which was originally selected for the Infrared Space Observatory (ISO) due to its exceptionally low Galactic infrared "cirrus" contamination \citep{oliver2000european}. Consequently, ELAIS-N1 has become a "legacy" field, targeted by numerous multi-wavelength campaigns to build a comprehensive census of galaxy evolution, including deep infrared surveys like the Spitzer Wide-Area Infrared Extragalactic Survey (SWIRE; \cite{Lonsdale_2003}) and deep radio observations from LOFAR \citep{Sabater_2021} and uGMRT (400 MHz; \cite{Arnab2019_2} and 1250 MHz; \cite{Sinha2023spectral}. Complementing this vast dataset is the UVIT aboard AstroSat, India's first dedicated astronomical observatory \citep{singh2014astrosat}. UVIT features twin 38-cm Ritchey-Chrétien telescopes that simultaneously image a $\sim 28$-arcminute field in three channels: Far-Ultraviolet (FUV, 130–180 nm), Near-Ultraviolet (NUV, 200–300 nm), and Visible (VIS, 320–550 nm) \citep{kumar2012ultra}. Its primary technical advantage is its high angular resolution ($\sim 1.2 - 1.8$ arcsec FWHM), which is significantly better than previous UV missions like GALEX. This resolution is achieved by using the VIS channel to track and correct for spacecraft jitter in post-processing, enabling its photon-counting FUV/NUV detectors to produce sharp images \citep{Tandon_2017, tandon2020orbit}.

The primary sources of ultraviolet (UV) emission in extragalactic systems are the thermal radiation from the hot, luminous photospheres of massive, young O- and B-type stars and the quasi-thermal continuum from the hot accretion disk surrounding a supermassive black hole (SMBH) in an AGN \citep{shakura1973black}. Evolved, low-mass populations like post-asymptotic giant branch stars also contribute, creating the "UV upturn" in elliptical galaxies \citep{oconnell1999hot}, but the flux in star-forming systems is dominated by the young, massive stars. This makes the UV continuum a powerful, direct tracer of instantaneous star formation on $\sim 10-100$ Myr timescales \citep{kennicutt1998star}. However, relying on UV data alone is insufficient due to two fundamental challenges: dust attenuation and source degeneracy. The interstellar medium (ISM) is highly opaque to UV photons, which are efficiently absorbed by dust grains and re-radiated thermally at mid- and far-infrared (IR) wavelengths \citep{calzetti2000dust}. Furthermore, the UV continuum from a starburst can be degenerate with that of a low-luminosity or obscured AGN. Multiwavelength data is essential to resolve these ambiguities. By combining the UV flux (tracing unobscured star formation) with the IR flux (tracing the obscured, re-radiated component), one can measure the total, dust-corrected star-formation rate (SFR) \citep{hao2011dust, kennicutt2012star}. More powerfully, assembling a panchromatic SED from IR, optical, and UV data allows for the physical decomposition of the observed flux. Different components dominate at distinct wavelengths: mid-IR colors can separate the hot, power-law continuum of an AGN's dusty torus from the cooler, Polycyclic Aromatic Hydrocarbon (PAH) feature rich dust heated by star formation \citep{donley2012identifying}; optical emission lines can classify sources via diagnostics like the BPT diagram \citep{baldwin1981classification}; and deep radio data traces both AGN jets and synchrotron emission from supernova remnants \citep{condon1992radio}. Fitting this comprehensive SED with composite models [e.g., \cite{conroy2013modeling, walcher2011fitting}] is the only robust method to break the degeneracies and simultaneously quantify the true SFR, stellar mass ($M_*$), and AGN luminosity.

A key principle to interpret the star formation activity of galaxies across cosmic time is the SFMS \citep{noeske2007star}. It is a tight, nearly linear correlation between the SFR and stellar mass of SFGs that has been observed observed in local universe, out to $z \sim 6$ \citep{speagle2014highly, schreiber2015herschel}. The existence of the the SFMS implies that the majority of the star formation proceeds through a steady, self-regulated mode of gas accretion and consumption, rather than violent, merger-driven starburst events \citep{elbaz2011goods, rodighiero2011lesser}. This secular evolution is mirrored in the star formation rate density, which increases till $z \sim 2-3$, and then declines exponentially to the present epoch \citep{Madau_2014}. The redshift range probed in this work, $0 < z \lesssim 0.76$, samples the latter $\sim 6-7$ Gyr of this decline. 

A galaxy's SED is a fundamental observable, representing a plot of its energy output as a function of wavelength, compiled from panchromatic photometric data. The resultant shape of the SED constitutes an integrated record of all physical processes occurring within the system. The UV-to-optical continuum is dominated by stellar emission; the UV component traces hot, massive, young stellar populations, whereas the optical continuum is primarily from the older, lower-mass stars that comprise the bulk of a galaxy's stellar mass \citep{bruzual2003stellar}. Diagnostic features in this range, such as the $4000$ \AA break ($D_n(4000)$), serve as a chronometer, revealing the mean age of the constituent stellar populations \citep{kauffmann2003stellar}. This stellar radiation is, however, subject to significant attenuation by the ISM. Interstellar dust grains absorb short-wavelength photons and subsequently re-radiate this energy in the IR. This re-processed emission spectrum contains its own diagnostic features, including mid-IR emission bands from PAHs and a broad, thermal continuum in the far-IR that is proportional to the total dust-obscured SFR \citep{calzetti2000dust, draine2007infrared} . The primary objective of SED fitting, therefore, is to deconvolve this complex, composite signal. By applying a rigorous energy balance principle, wherein the energy attenuated by dust in the UV/optical regime must be precisely equivalent to the energy re-radiated in the IR, codes such as CIGALE (Code Investigating GALaxy Emission) can model these components simultaneously. This methodology permits the derivation of robust physical properties, including the intrinsic star-formation rate, total stellar mass, and the fractional contribution of any AGN \citep{noll2009analysis, boquien2019cigale}.

In the UV domain, extensive imaging has been utilsed to investigate the properties of unobscured star formation. \cite{pasquali2006morphology} performed a detailed morphological analysis of UV-detected galaxies in the ELAIS-N1 field using Galaxy Evolution Explorer (GALEX) data. This study classified galaxies based on their UV structures (e.g., clumpy, spiral, smooth) and demonstrated how these UV morphologies correlate with the physical properties of the stellar populations, linking the structural appearance directly to the mode and intensity of recent star formation.

A deep FUV photometric survey exploiting the angular resolution of AstroSat/UVIT, combined with panchromatic SED derived physical propoerties across a continous redshift baseline of $0 < z \lesssim 0.76$, has not previously been presented for this field. Prior UV studies of ELAIS N1, such as the morphological analysis of \citep{pasquali2006morphology}, relied on GALEX data, which is limited by coarse angular resolution of $\sim 5''$ FHWM. This resolution is insufficient to deblend close sources in a deep field at cosmological distances, leading to source confusion and flux contamination that biases the photometric measurements and consequently, derived physical properties. The UVIT with its $\sim 1.5''$ PSF, represents an improvement of more than a factor of three in angular resolution over GALEX. The availability of the highly accurate redshift catalogue of \cite{Duncan_2021} provides the redshift baseline required to study the physical properties as a function of time. In this work, we exploit these complementary datasets to construct a robust FUV source catalogue perform multiwavelength SED fitting using CIGALE \citep{boquien2019cigale}, and derive SFRs, stellar mass, and young stellar mass fractions as a function of redshift for a sample of SFGs.

The paper is divided into following sections; Section \ref{sec:Obserations} describes the UVIT observation and data reduction, PSF modelling, background estimation, sources of ancillary optical and IR data, and photometric redshift sources; Section \ref{sec:cat} describes the method of catalogue preparation and AGN classification; Section \ref{sec:sed} describes the SED modelling; Section \ref{sec:props} talks about the how SFR and stellar mass vary with redshift; Section \ref{sec:conc} presents the conclusions of this work. Throughout this work, we assume a flat $\Lambda$CDM cosmology with $H_0 = 70 \ km s^{-1} \ Mpc^{-1}$, $\Omega_m = 0.3$, and $\Omega_{\Lambda} = 0.7$, and all magnitudes are quoted in the AB system.

\section{Observations and Data Analysis} \label{sec:Obserations}

We used the UVIT instrument onboard the \textit{AstroSat} satellite to observe the ELAIS\,N1 field at ultraviolet wavelengths. The UVIT instrument observes primarily in two bands: FUV (1300--1800\,\AA) and NUV (2000--3000\,\AA). In addition, it can also observe in the visible channels (3200--5500\,\AA), which are mainly used to track the drifting pattern of the satellite pointing while observing. The instrument is capable of simultaneous observation across all three wavelength bands, but the FUV and NUV wavebands are predominantly used for scientific observations. The detectors in these bands work in the photon counting mode with a frame rate of 28.7\,Hz. The field of view (FoV) is circular, with a diameter of nearly $28'$. We refer an interested reader to \citet{Tandon_2017} for details on the UVIT instrument and its in-orbit calibrations.

\begin{figure*}
    \centering
    \includegraphics[width=1\linewidth]{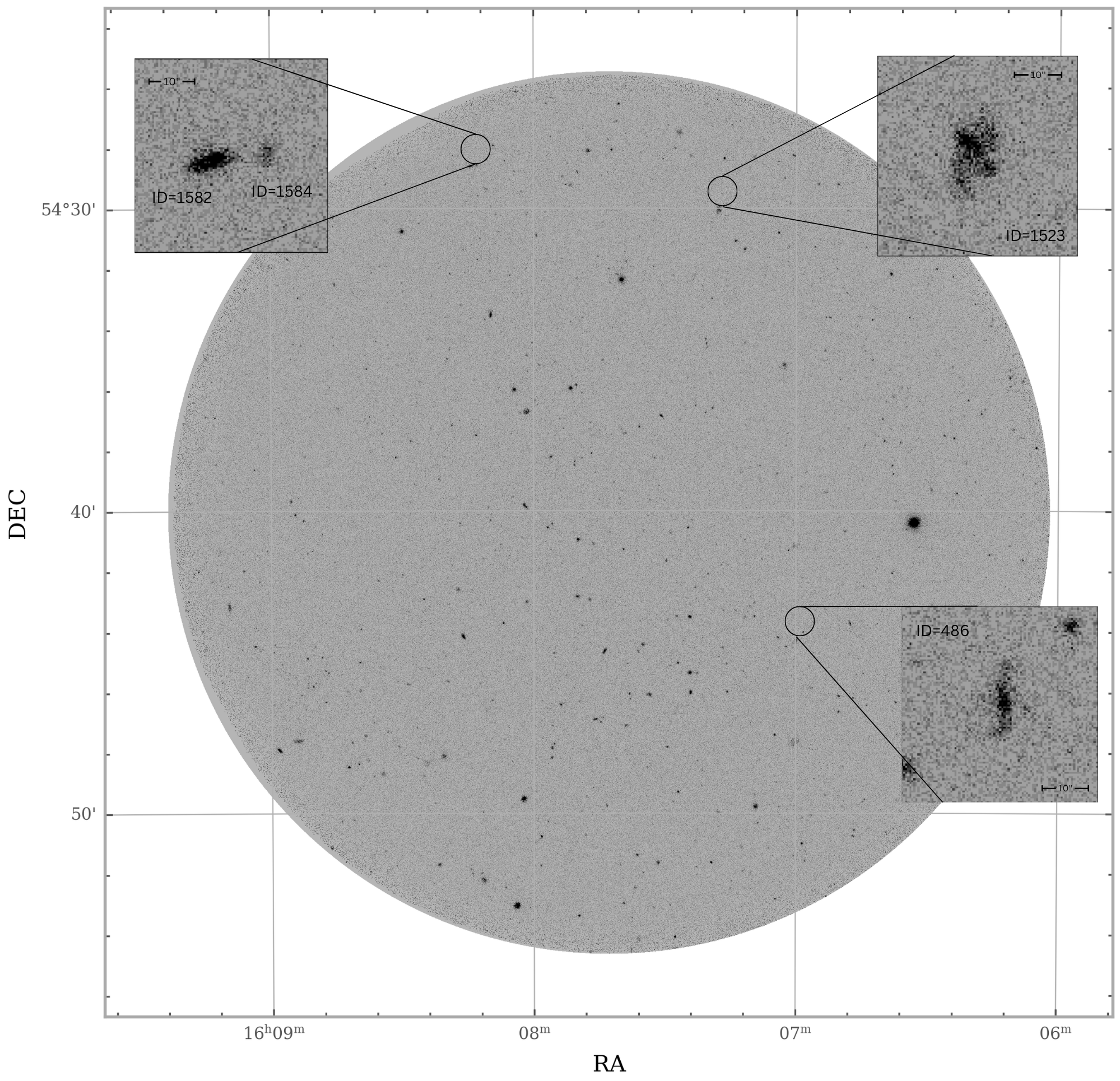}
    \caption{ELAIS N1 field observed by AstroSat/UVIT in FUV using F154W filter with some sources selected based on visual morphology overlaid. The sources are described in sec \ref{sec:Obserations}.}
    \label{fig:EN1_UVIT_image}
\end{figure*}

We observed the ELAIS\,N1 field centred at RA$= 16^{\rm h} 07^{\rm m} 33^{\rm s}$, DEC$=54^{\rm d} 38^{\rm m} 31^{\rm s}$ at FUV band using the UVIT (Proposal-code: A09\_013). The broad-band filter BaF2 ($\lambda_{\rm eff} = 1541$\,\AA, F154W) was used for the observations with a total on-source exposure time of 30 ksec. The Level 1 data was obtained from the Indian Space Science Data Center (ISSDC/ISRO). The Level 2 (science ready) images were produced using CCDLAB v3.0 \citep{postma2017ccdlab, postma2021ccdlabtutorial}. The UVIT Calibration Database \citep{girish2017mapping, postma2011calibration} contains information about the necessary systemic corrections due to imperfections in detectors, such as Fixed Pattern Noise (FPN), and Camera Proximity Unit (CPU) Distortion. Then the frames affected by cosmic rays are removed, although there are not many frames that are affected by cosmic rays. Then the images are scaled with the exposure map to normalise the image in order to get a uniform exposure time. Additionally, the exposure map removes the "bad pixels" as found in ground calibration. Next, the images are flat fielded to remove the remaining geometrical defects. After this, drift correction is applied using VIS data. The orbitwise images are then displayed and we then select 3 bright sources visible across all images to get a merged image. We optimised the PSF to eliminate the remaining drift that had been missed earlier. The resulting image is displayed in Figure \ref{fig:EN1_UVIT_image}. The overlaid sources were selected based on their visual morphologies. 
\begin{enumerate}
    \item Source ID 1582 and 1584 appear to be edge-on disk galaxy with z $\sim$ 0.12 with a smaller spherical source $\sim$ 64 Mpc away.
    \item Source ID 1523 is a source at z $\sim$ 0.07; the source exhibits a complex structure.
    \item Source ID 486 is a elongated, slightly asymmetric structure at z $\sim$ 0.09.
\end{enumerate}

In the following sections, we derive the Point Spread Function (PSF), background, source of multiwavelength data and redshifts.

\begin{table}
    \centering
     \caption{Observation Summary of the ELAIS N1 field using the UVIT instrument.}
    \begin{tabular}{c|c}
    \hline \hline
    Project code &  A09\_013 \\
      Band       &  FUV \\
      Filter   &    F154W \\
      $\lambda_{\rm eff}$   &   1541\AA \\
      ZP magnitude  &   17.778 \\
      Unit Conversion   &   3.55$\times10^{-15}$ \\
                    & ($\rm{erg}\,\rm{sec^{-1}}\,\rm{cm^{-2}}\,$\AA$^{-1}$) \\
      Exposure time     &   30,000 sec\\
      \hline
    \end{tabular}
   
    \label{tab:my_label}
\end{table}

\subsection{Point Spread Function}\label{sec:psf} 

We discuss the methods to measure robust estimates of the Full Width Half Maxima (FWHM) of the Point Spread Function (PSF) and the background of the FUV map in this section. 

We select stars in the field by matching with the GAIA DR3 catalogue \citep{prusti2016gaia, vallenari2023gaia}. The cutouts of the stars are median stacked to get the PSF. The median stacked image is then visually inspected to check the symmmetry around the centroid. Then we use the \texttt{photutils.profiles} \citep{larry_bradley_2024_13989456} to create the radial profile and curve of growth of the median stacked image. The radius at the 80\% flux was found to be $9.14$ arcsec ($3.81$ pixels).

We use the circular Moffat function \citep{Moffat_1969} to fit the surface brightness distribution, $I(r)$ as a function of radius to measure the FWHM. Since the wings of a stellar profile cannot be effectively fitted by a Gaussian function alone \citep{Trujillo_2001}, we employ the Moffat function in this case for simplicity. For information on wing modelling in the BaF2 filter, see \cite{Kanak_2021}. A restricted version of the Moffat function is the Gaussian function, i.e., when $\beta \to \infty$. 

The Moffat function is given as,
\begin{equation}
    I(r) = I_0\left[ 1+\left(\frac{r}{\alpha}\right)^2 \right]^{-\beta}
\end{equation}
where $I_0$ is the central surface brightness with $\alpha$ and $\beta$ as free parameters. $\beta$ determines the spread of the function while the FWHM is given as, FWHM = $2\alpha\sqrt{2^{(1/\beta)}-1}$. The fitting gives the values for the PSF FWHM of 3.61 pixels ($1''.5$) for the F154W filter.

\begin{figure}
    \centering
    \includegraphics[width=\columnwidth]{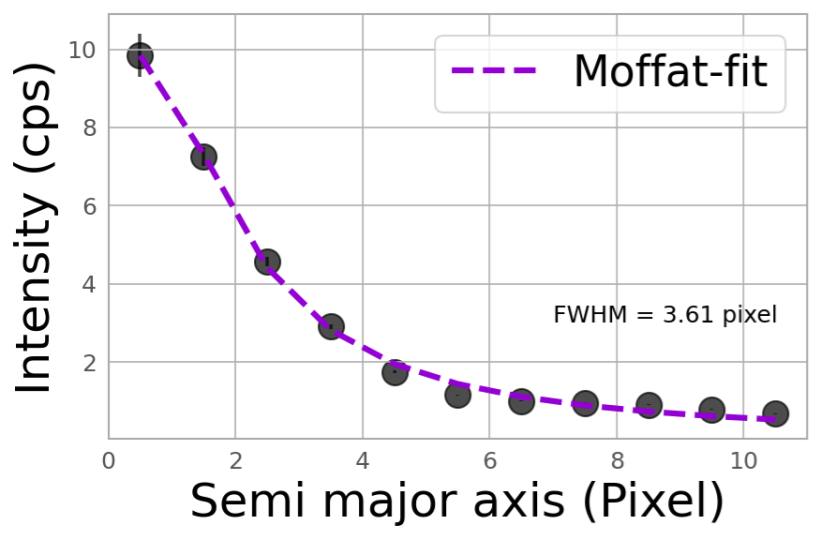}
    \caption{The PSF obtained by fitting the Moffat function using its surface brightness distribution to measure the FWHM (see section \ref{sec:psf}).  }
    \label{fig:psf}
\end{figure}

\subsection{Background Estimation}\label{bkg}

For robust estimation of source flux and the signal-to-noise ratio (SNR), particularly the faint ones, it is essential to comprehend the sky background in the photometric study of the deep fields. Here, we analyse the FUV background from the deep UVIT observations in detail. We use \texttt{photutils.background} python package to estimate the background. 

We use a box size of $25 \times 25$ pixels to create a grid. We use $3 \sigma$ clipping to distinguish the bright sources from the background. To estimate the background in the individual boxes, we use the \texttt{MedianBackground()} estimator. To avoid issues with some large sources, we use a filter of $3 \times 3$ pixels to suppress the affected pixels.

We get an RMS background of $1.38 \times 10^{-5}$ cps. We get a $3\sigma$ limiting magnitude of $25.69 \ m_{AB}$ and a $5 \sigma$ magnitude of $25.13 \ m_{AB}$.

\subsection{Other Multi-wavelength Data}

\subsubsection{Optical}\label{sec:optical}

We have used the third phase dark time survey of the Sloan Digital Sky Survey \citep[SDSS-III]{York_2000}, also called the Baryon Oscillation Spectroscopic Survey (BOSS). This survey observed 1.5 million massive galaxies to measure the distance–redshift relation $d_A(z)$ and the Hubble parameter $H(z)$ with per\,cent-level precision out to z = 0.7 \citep{Eisenstein2005,Cole_2005} over an area of 10,000\,deg$^2$. A series of plates were granted to observe and obtain spectra for the sources in the ELAIS N1 observed in radio wavelengths using GMRT, JVLA, LOFAR and FIRST. These catalogues were made public as a part of the SDSS Data Release (DR) 12. 
We use these spectroscopic redshifts ($z_{\rm spec}$) for our UV sources in this work. Besides, the spectroscopic redshift catalogue also provides the source classification based on the spectroscopy, the details of which are described in \citet{Bolton_2012}. 
Furthermore, we use the 16th DR of the SDSS catalogue provided by \cite{Ahumada_2020} to obtain the $u$,$g$,$r$,$i$ and $z$ magnitudes of our UV sample. 

\subsubsection{Infrared}

The ELAIS N1 is one of the six extragalactic deep fields that were observed using the \textit{Spitzer} telescope in the widely known survey called the SWIRE survey \citep{Rowan_2008,Rowan-Robinson2013}. This survey covered an area of 49$\,\deg^2$ of which the ELAIS N1 field was observed in a sky area of 8.72\,$\deg^2$ using the Infrared Array Camera (IRAC) at 3.6, 4.5, 5.8 \& 8\,$\upmu$m and with the Multi-Band Imaging Photometer (MIPS) at 24, 70 \& 160\,$\upmu$m \citep{Lonsdale_2003, Mauduit_2012}. In fact, this survey and the revised catalogue also include other wavebands: the five bands, $U',g',r',i'$ and $Z'$ using the Wide Field Survey using the 2.5-m Isaac Newton Telescope \citep{McMahon_2001}, the $J,H$ and $K_{s}$ bands using the Two Micron All Sky Survey (2MASS) and he near-infrared from the UKIRT Infrared Deep SkySurvey \citep[UKIDSS;][]{Lawrence_2007}. The SWIRE catalogue is thus one of the reliable catalogues with the availability of huge photometric data. 

\subsection{Photometric Redshift}

We are using photometric redshifts provided by \cite{Duncan_2021} for ELAIS N1 deep field. They estimate photometric redshifts using a hybrid methodology that combines traditional template-fitting with machine learning techniques within a Hierarchical Bayesian (HB) framework. This produces robust consensus redshifts optimised for the diverse populations of star-forming galaxies and AGN found in deep radio surveys. This method is a significant evolution from the technique used for the shallower LoTSS First Data Release (DR1) \citep{duncan2019lofar}, which was foundational but less refined. The key improvement in the 2021 work was the optimization of the covariance hyperparameter, $\beta$, making it a function of source magnitude rather than a single constant value. This refinement was critical for accurately quantifying redshift uncertainties across the wider dynamic range of the deeper survey, where the previous approach underestimated errors for the brightest and faintest sources. The result is a more precise redshift catalogue, with scatter for galaxies improving from $\sim ~3.9\%$ in DR1 to $\sim 1.6-2.0\%$ in the Deep Fields. We get photometric redshift for 1196 sources at $3\sigma$ detection and 422 sources at $5\sigma$ detection. The Normalised Median Absolute Deviation ($\sigma_{\mathrm{NMAD}}$) is 0.2, with an outlier fraction of 0.16. The estimates have scatter $< 0.04 \times (1+z)$ for galaxy dominated population at $z<1$.
Additionally, we also use the spectroscopic redshifts given by BOSS \citep{abdurro2022seventeenth}. It is available for only a small fraction of the sample, 51 at $3\sigma$ detection and 30 at $5\sigma$ detection. Preference is given to spectroscopic redshift and photometric redshift is used for the samples where spectroscopic redshift is not available. The pipeline statistical uncertainties translate to $\Delta z = 10^{-4}$.

\section{Source Catalogue and Classification} \label{sec:cat}

We use Source Extractor Package \citep{barbary2016sep}, which makes the crucial features of Source Extractor \citep[SExtractor;][]{Bertin_1996} in python, to detect sources in the FUV image. We created the curve of growth of the point sources using \texttt{photutils.profiles} \citep{larry_bradley_2024_13989456} and found the 80\% of the flux was contained in a radius of 9.14 pixels. Thus, we used a default convolution filter with a \texttt{PHOT\_APERTURES} diameter of 9.14 pixels. We use 32 deblending-subthreshold with 0.005 as the minimum contrast parameter and \texttt{CLEAN\_PARAM} = 1.0. Further, a pixel-scale value of 0.416 was used and \texttt{MAG\_ZEROPOINT} = 17.778 to obtain the FUV magnitudes in AB magnitudes.

To accurately measure the intrinsic luminosities of the galaxies, particularly in the highly sensitive FUV band, it is imperative to correct the observed photometry for foreground Galactic extinction. We retrieved the line-of-sight Galactic color excess, $E(B-V)$, for the ELAIS N1 field coordinates using the Galactic Dust Reddening and Extinction tool hosted by the NASA/IPAC Infrared Science Archive (IRSA) \citep{https://doi.org/10.26132/ned5}. This tool derives reddening estimates based on high-resolution, all-sky maps of Galactic dust thermal emission \citep{schlegel1998maps, schlafly2011measuring}. Because the ELAIS N1 region was originally selected for deep extragalactic surveys specifically due to its low infrared cirrus background, the foreground reddening in this direction is minimal. Nevertheless, the position-dependent $E(B-V)$ values were extracted and converted into specific extinction values, $A_\lambda$, for the AstroSat/UVIT F154W filter by calculating $k'(\lambda)$ given by \cite{calzetti2000dust}. The observed magnitudes were subsequently dereddened ($m_{\mathrm{intrinsic}} = m_{\mathrm{obs}} - A_\lambda$) prior to performing the multi-wavelength SED fitting, ensuring that the derived star formation rates and stellar masses are not artificially suppressed by our own galaxy's interstellar medium.

\begin{figure}
    \centering
    \includegraphics[width=1.0\linewidth]{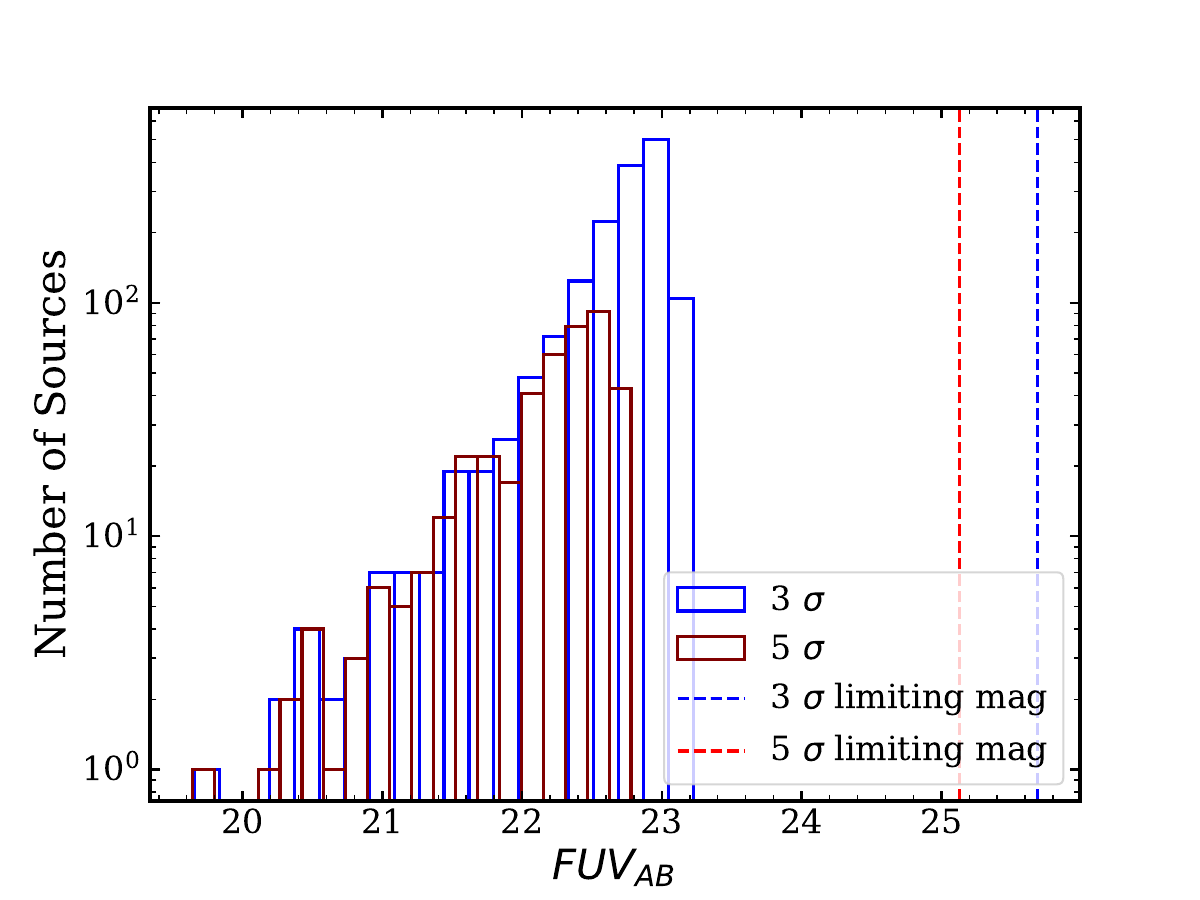}
    \caption{Distribution of the dereddened F154W magnitude}
    \label{fig:mag_dist}
\end{figure}

The distribution of the dereddened F154W magnitudes at $5\sigma$ and $3\sigma$ are shown in Figure \ref{fig:mag_dist} in red and blue histograms respectively.
We would like to mention that henceforth we will consider only the $3\sigma$ catalogue for further analyses in this work.

\subsection{Catalogue Description}
We get 1637 objects at $3 \sigma$ detection threshold and 458 objects at $5 \sigma$ detection threshold. Table \ref{tab:cat_desc} describes the columns in the catalogue for both, $3 \ \sigma$ source catalogue and $5 \ \sigma$ source catalogue. Table \ref{tab:cat_samp} shows the first 10 rows of the 3$\sigma$ catalogue.

\begin{figure}
    \centering
    \includegraphics[width = 1.0\linewidth]{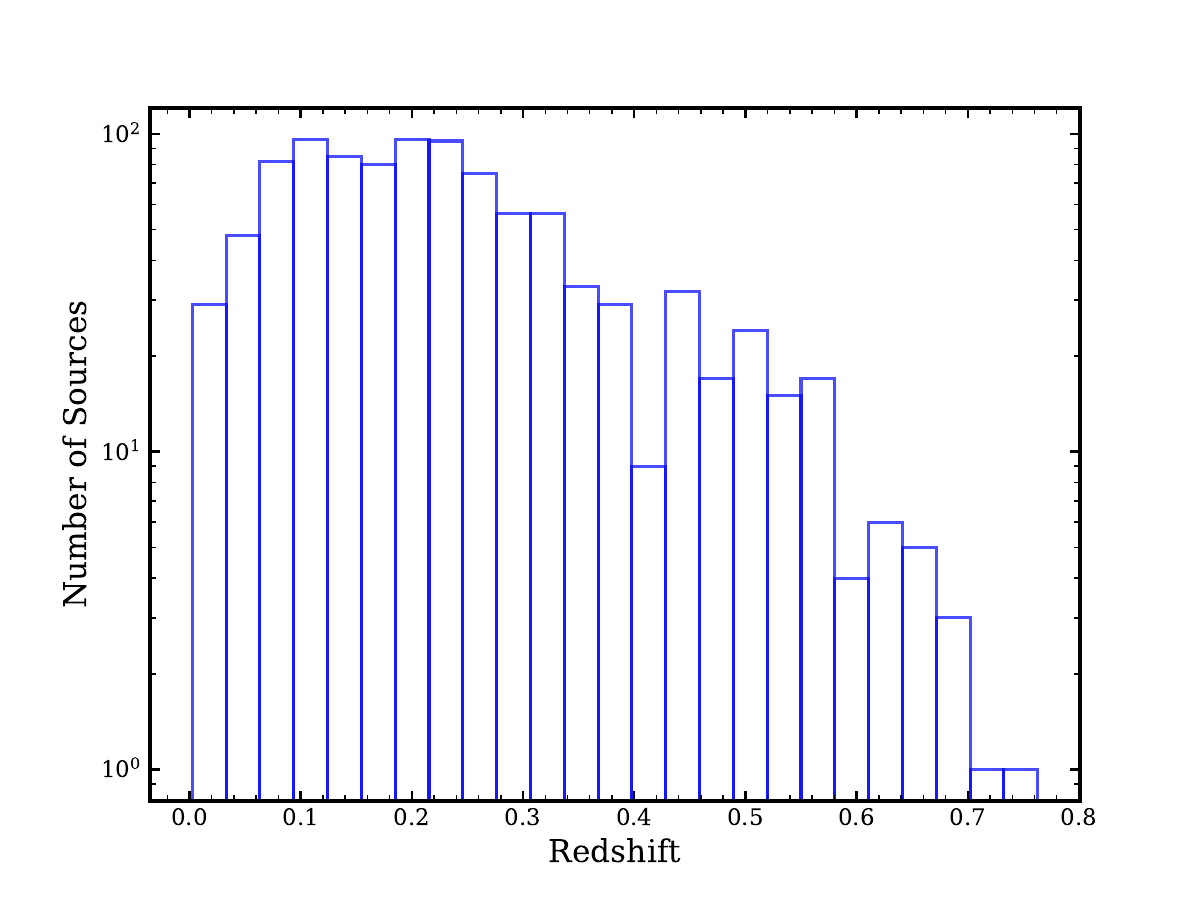}
    \caption{Redshift distribution of the SFGs detected at 3 $\sigma$ threshold.}
    \label{fig:redshift_dist}
\end{figure}

Figure \ref{fig:redshift_dist} shows the distribution of the redshift of the sources. The sources have a median redshift of 0.24. 

\begin{table}
    \centering
    \caption{Column description of the FUV catalogue}
    \begin{tabular}{c|c} \hline
        Column Name & Description \\ \hline
        Object$\_$ID & Identification number in catalogue\\
        x$\_$pixel & X pixel coordinate of the centroid \\
        y$\_$pixel & Y pixel coordinate of the centroid \\
        RA & RA of centroid in J2000 coordinate system \\
        DEC & DEC of centroid in J2000 coordinate system \\
        FUV$\_$mag & Magnitude within an aperture of radius $3.8"$ \\
        FUV$\_$mag$\_$err & Error in magnitude \\
        FUV$\_$kron$\_$mag & Magnitude within kron aperture \\
        FUV$\_$kron$\_$mag$\_$err & error in kron magnitude \\
        a$\_$image & Semi-major axis of the object (in pixels) \\
        b$\_$image & Semi-minor axis of the object (in pixels) \\
        theta & Position angle of the object (in degrees) \\ \hline
    \end{tabular}
    \label{tab:cat_desc}
\end{table}

\begin{table*}
    \centering
    \caption{First 10 rows of the 3$\sigma$ catalogue. The complete table is provided as an online table.}
    \resizebox{\textwidth}{!}{
    \begin{tabular}{c|c|c|c|c|c|c|c|c|c|c|c} \hline
        Object\_ID & x\_pixel & y\_pixel & RA & DEC & FUV\_mag & FUV\_mag\_err & FUV\_kron\_mag & FUV\_kron\_mag\_err & a\_image & b\_image & theta \\ \hline
        1 & 2258.68  & 483.12 & 241.96 & 54.89 & 20.99 & 24.77 & 21 & 26.1 & 2.96 & 1.93 & 0.002   \\
        2 & 2026.69 & 521.55 & 242 & 54.89 & 22.55 & 25.44 & 21.78 & 25.74 & 2.67 & 1.66 & -0.2494 \\
        3 & 2611.51 & 521.52 & 241.89 & 54.89 & 22.47 & 25.41 & 21.51 & 25.65 & 2.89 & 1.87 & 1.409   \\
        4 & 2614.66 & 521.67 & 241.89 & 54.89 & 22.42 & 25.39 & 21.9 & 25.91 & 2.71 & 1.27 & -1.2657 \\
        5 & 2111.83 & 526.96 & 241.99 & 54.88 & 22.93 & 25.57 & 22.6 & 26.22 & 1.74 & 1.04 & 1.2653  \\
        6 & 2741.6 & 528.49 & 241.86 & 54.88 & 22.93 & 25.57 & 22.95 & 26.37 & 1.6 & 0.94 & 1.181   \\
        7 & 2803.69 & 533.79 & 241.85 & 54.88 & 22.81 & 25.53 & 22.43 & 26.13 & 1.86 & 1.19 & -1.4811 \\
        8 & 2690.48 & 531.97 & 241.87 & 54.88 & 22.49 & 25.42 & 21.56 & 25.67 & 3.17 & 1.84 & -1.0082 \\
        9 & 2472.76 & 540.08 & 241.91 & 54.88 & 21.76 & 25.12 & 20.71 & 25.26 & 4.42 & 2.88 & 0.288   \\
        10 & 2469.09 & 541.53 & 241.91 & 54.88 & 21.77 & 25.12 & 20.98 & 25.48 & 3.19 & 2.8 & 0.3206  \\
        ...  & & & & & & & & & & & \\
        \hline
    \end{tabular}
    }
    \label{tab:cat_samp}
\end{table*}

\subsection{AGN Classification}
UV emissions typically emanate from star-forming regions within host galaxies. However, in the case of certain sources like active galactic nuclei (AGN), a distinctive feature known as the `Big Blue Bump` (BBB) appears in their spectral energy distribution, where emission peaks in the UV range \citep{sanders1989continuum}. Consequently, it’s crucial to differentiate between SFGs and AGN within the FUV sources using multi-wavelength criteria.

We use the AGN classification given by \cite{Duncan_2021}. They have used 3 criteria for classifying a source as AGN:
\begin{enumerate}
    \item If the source is in the Half Million Quasar Catalogue \citep{Flesch_2015}
    \item If the source fulfils \cite{donley2012identifying} selection criteria. We get 451 objects after cross-matching the UVIT catalogue with the IRAC catalogue.  Due to the depth of our survey, we detect a large number of faint objects, and as this method is inherently biased towards bright AGNs and host-dominated AGNs, we get only 170 sources with observations in all 4 channels in IRAC. 266 sources are missing observations in the 5.8-micron channel, 183 sources are missing observations in the 8.0-micron channel, and 168 sources are missing observations in both channels.
    \item If the source is detected in the Second ROSAT All-Sky Survey (2RXS) \citep{Boller_2016} and in the XMM-Newton Slew Survey (XMMSL2) \footnote{\url{https://www.cosmos.esa.int/web/xmm-newton/xmmsl2-ug\#}}. Both these surveys have observations around the observed field, but there are no observations in the FoV.
\end{enumerate}

If a source fulfils any of the above conditions, it is classified as an AGN.

In our sample, we get 23 sources classified as AGNs with these criteria.

\section{Multiwavelength Spectral Energy Distribution (SED)} \label{sec:sed}

CIGALE \citep{boquien2019cigale} is a powerful tool widely used for generating SEDs. It allows a wide range of input parameters, such as star formation history, stellar properties, nebular emission, dust attenuation, AGN, etc. It compares modelled galaxy SEDs generated using all the possible combinations of the input parameters with the observations provided. Values for all the parameters are given in the table \ref{tab:cigale_summary}.

For detailed information on the modules, we ask the reader to check the descriptions given in \cite{boquien2019cigale}.

\begin{figure}
    \centering
    \includegraphics[width=1.0\linewidth]{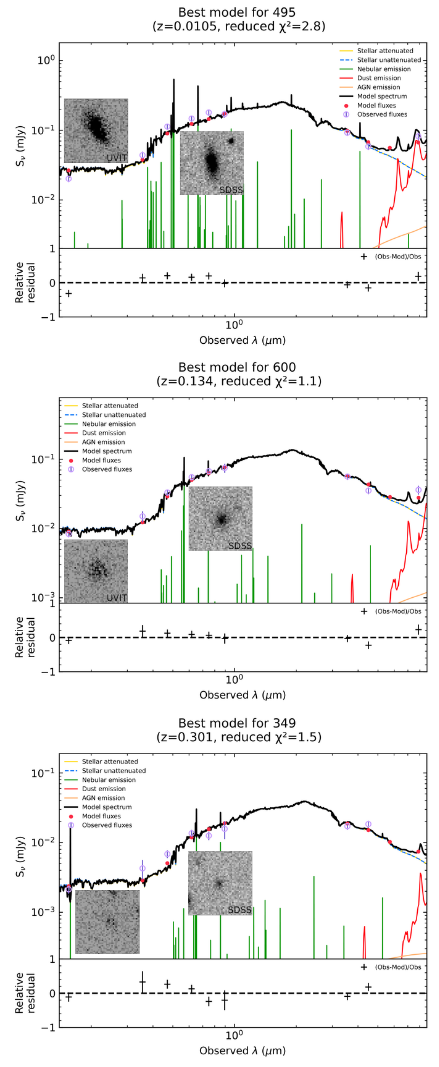}
    \caption{SED fits of a few selected sources with their cutouts in the FUV band from UVIT and the r band from SDSS; the cutouts are 20 arcsec on both axes.}
    \label{fig:sed}
\end{figure}

Figure \ref{fig:sed} displays the representative multiwavelength SED fitting outputs generated using CIGALE, paired with corresponding $20^" \times 20^"$ image cutouts. The left panels shows the best-fit SED models for three representative galaxies at varying redshifts. These plots compare observed broadband fluxes (purple circles) and model fluxes (red dots) against the total model spectrum (black line), which consists of unattenuated and attenuated stellar, nebular, dust, and AGN emission components. Relative residuals plotted below each SED confirm the robustness of the fits through low reduced $\chi^2$ values. The right panels visually present the spatial morphology of these respective sources in both the ultraviolet (FUV from UVIT) and optical (r band from SDSS) regimes.

\begin{table*}
    \centering
    \caption{Input parameters for CIGALE}
    \resizebox{\textwidth}{!}{
    \begin{tabular}{c|c|c}\toprule
    Module &Parameter &Values \\\midrule
    \multirow{6}{*}{Delayed Star Formation History with an additional burst: [sfhdelayed]} &e-folding time of the main stellar population (Myr) &2000.0, 4000.0, 6000.0, 8000.0 \\
    &Age of the main stellar population (Myr) &6000.0, 8000, 10000, 12000 \\
    &e-folding time of the late starburst population (Myr) &50.0, 100.0 \\
    &Age of the late burst population (Myr) &250, 500, 750 \\
    &Mass fraction of late burst population &0.0, 0.005, 0.01 \\
    &Normalise the SFR to 1 solar mass &True \\ \hline
    \multirow{3}{*}{Single stellar population: [bc03]} &Initial Mass Function &\cite{chabrier2003galactic} \\
    &metallicity &0.02, 0.05 \\
    &Age separation of young and old stars (Myr) &10 \\ \hline
    \multirow{7}{*}{Nebular emission: [nebular]} &Ionisation parameter &-3.0, -2.0, -1.0 \\
    &Metallicity of gas &0.011, 0.02, 0.03 \\
    &Electron density &100 \\
    &Escape fraction of Lyman continuum photons &0.1, 0.4 \\
    &Fraction of Lyman continuum photons absorbed by dust &0.1, 0.4 \\
    &Line width (km/s) &300 \\
    &Include nebular emission &True \\ \hline
    \multirow{6}{*}{Attenuation law: [dustatt\_calzleit]} &Color excess of the stellar continuum light for young stars &0.005 \\
    &Color excess of the stellar continuum light for old stars &0.3, 0.1 \\
    &Central wavelength of the UV bump (nm) &217.5 \\
    &FWHM of the UV bump (nm) &35 \\
    &Amplitude of the UV bump &0.0, 2.0 \\
    &Slope of the attenuation curve power law &-0.3, -0.1, 0 \\ \hline
    \multirow{2}{*}{Dust emission: [dale2014]} &AGN fraction &0.02 \\
    &Slope &1.5, 2.5 \\ \hline
    \multirow{8}{*}{Active galactic nuclei: [skirtor2016]} &Average edge-on optical depth at 9.7 micron &5, 7, 9 \\
    &Power-law exponent of the radial gradient of dust density &0.5, 1.5 \\
    &Index of dust density with polar angle &0.5, 1.5 \\
    &Angle measured between equatorial and edge of the torus &40, 60 \\
    &Ratio of outer to inner radius &20 \\
    &fraction of dust mass inside clumps &0.97 \\
    &Viewing angle &30 \\
    &AGN fraction &0.05 \\
    \bottomrule
    \end{tabular}
    }
    \label{tab:cigale_summary}
\end{table*}

\subsection{Star Formation History}

We choose a delayed star formation model with an optional late burst called \texttt{sfhdelayed}. This model has a smoother variation with time. This provides a flexible and physically viable analytical model. The delayed SFR is of the form $SFR(t) \propto \frac{t}{\tau^2}exp(-t/\tau)$ when $0 \leq t \leq t_0$ where $t_0$ is the age of the universe at the time redshift of the galaxy, $\tau$ is the age at the peak of star formation. The SFR steadily increases with time and decreases after reaching the peak. We use a large value of $\tau$ to get a late peak.

\section{Physical Properties of Sources} \label{sec:props}

Here we will discuss the statistical properties of galaxies, like SFR, mass of stars, and mass of young stars in sources.The statistics of the derived physical properties are displayed in Table \ref{tab:physicalprops}. The mean and the standard deviation of the SED parameters are shown in Table \ref{tab:sed_parameters}.
The summary of the physical properties is as follows:

\begin{table*}
    \centering
    \caption{Summary of the physical properties. z$_{\mathrm{range}}$ is the redshift bin; z$_{\mathrm{mean}}$ is the mean of the redshift bin; N$_{\mathrm{sources}}$ is the number of sources in the redshift bin; 'SFR' is the median SFR in the redshift bin; log(M$_{\mathrm{star}}$) is the log of the median of the stellar mass in the redshift bin; log(M$_{\mathrm{young}}$) is the log of the median of the young stellar mass in the redshift bin; and log(M$_{\mathrm{young}}$/M$_{\mathrm{star}}$) is the log of the median of the  ratio of young stellar mass to total stellar mass in the redshift bin.}
    \begin{tabular}{c|c|c|c|c|c|c} \hline
         z$_{\mathrm{range}}$ & z$_{\mathrm{mean}}$ & N$_{\mathrm{sources}}$ & SFR $[M_{\odot} yr^{-1}]$ & log(M$_{\mathrm{star}} [M_{\odot}]$) & log(M$_{\mathrm{young}} [M_{\odot}$]) & log(M$_{\mathrm{young}}$/M$_{\mathrm{star}}$)\\ \hline
         
         0.002-0.1 & 0.04 & 20 & 0.02 $\pm$ 0.00 & 8.25 $\pm$ 7.38 & 5.27 $\pm$ 4.12 & $(112 \pm 5.55) \times 10^{-5}$\\
         
         0.1-0.2 & 0.16 & 28 & 0.39 $\pm$ 0.01 & 9.66 $\pm$ 8.26 & 6.58 $\pm$ 4.97 & $(72.3 \pm 4.65 )\times 10^{-5}$\\

         0.2-0.3 & 0.24 & 39 & 0.66 $\pm$ 0.02 & 9.98 $\pm$ 8.46 & 6.80 $\pm$ 5.18 & $(56.2 \pm 4.48) \times 10^{-5}$\\

         0.3-0.4 & 0.33 & 36 & 1.29 $\pm$ 0.03 & 10.23 $\pm$ 8.62 & 7.10 $\pm$ 5.42 & $(58.0 \pm 4.16) \times 10^{-5}$\\

         0.4-0.5 & 0.46 & 26 & 2.69 $\pm$ 0.07 & 10.31 $\pm$ 8.96 & 7.41 $\pm$ 5.83 & $(131 \pm 6.61) \times 10^{-5}$\\

         0.5-0.76 & 0.56 & 25 & 4.27 $\pm$ 0.10 & 10.67 $\pm$ 9.04 & 7.62 $\pm$ 6.01 & $(124 \pm 5.35) \times 10^{-5}$\\
         \hline
    \end{tabular}
    \label{tab:physicalprops}
\end{table*}

\begin{table*}
    \centering
    \caption{Mean and Standard Deviations of the SED parameters}
    \resizebox{\textwidth}{!}{
    \begin{tabular}{c|c|ccccccc} \hline
         Parameter &  & z$\leq$0.1 & 0.1$\leq$z$<$0.2 & 0.2$\leq$z$<$0.3 & 0.3$\leq$z$<$0.4 & 0.4$\leq$z$<$0.5 & z$>$0.5 & Full Sample \\ \hline
        \multirow{2}{*}{\begin{tabular}[c]{@{}c@{}}Age of main stellar \\ population\end{tabular}} & Mean & 8000 & 8285.714 & 8820.513 & 8111.111 & 7230.769 & 7120 & 8011.494 \\
        & SD & 2366.432 & 1665.986 & 1482.752 & 1328.696 & 973.009 & 992.774 & 1611.698 \\
        \multirow{2}{*}{\begin{tabular}[c]{@{}c@{}}e-folding time of main \\ stellar population\end{tabular}} & Mean & 3100 & 2571.429 & 3128.205 & 2666.667 & 3230.769 & 2960 & 2931.034 \\
        & SD & 1946.792 & 1293.626 & 1742.081 & 1154.701 & 1846.154 & 1708.333 & 1628.01 \\
        \multirow{2}{*}{\begin{tabular}[c]{@{}c@{}}Age of late starburst \\ population\end{tabular}} & Mean & 375 & 321.429 & 346.154 & 354.167 & 326.923 & 360 & 346.264 \\
        & SD & 167.705 & 131.223 & 156.232 & 170.528 & 134.615 & 159.374 & 155.365 \\
        \multirow{2}{*}{\begin{tabular}[c]{@{}c@{}}e-folding time of late \\ starburst population\end{tabular}} & Mean & 57.5 & 64.286 & 62.821 & 54.167 & 61.538 & 56 & 59.483 \\
        & SD & 17.854 & 22.588 & 21.833 & 13.819 & 21.066 & 16.248 & 19.601 \\
        \multirow{2}{*}{Fraction of late starburst} & Mean & 0.007 & 0.008 & 0.007 & 0.008 & 0.008 & 0.008 & 0.008 \\
        & SD & 0.004 & 0.004 & 0.004 & 0.004 & 0.004 & 0.004 & 0.004 \\
        \multirow{2}{*}{Stellar metallicity} & Mean & 0.023 & 0.026 & 0.035 & 0.035 & 0.026 & 0.028 & 0.03 \\
        & SD & 0.009 & 0.012 & 0.015 & 0.015 & 0.012 & 0.013 & 0.014 \\
        \multirow{2}{*}{Gas metallicity} & Mean & 0.015 & 0.016 & 0.016 & 0.015 & 0.014 & 0.014 & 0.015 \\
        & SD & 0.007 & 0.008 & 0.008 & 0.008 & 0.007 & 0.007 & 0.008 \\
        \multirow{2}{*}{\begin{tabular}[c]{@{}c@{}}Fraction of Luman continuum\\ photons absorbed by dust\end{tabular}} & Mean & 0.28 & 0.282 & 0.254 & 0.292 & 0.192 & 0.22 & 0.255 \\
        & SD & 0.147 & 0.147 & 0.15 & 0.144 & 0.138 & 0.147 & 0.15 \\
        \multirow{2}{*}{Dust alpha slope} & Mean & 1.95 & 2.107 & 2.013 & 2.167 & 2.077 & 2.18 & 2.086 \\
        & SD & 0.497 & 0.488 & 0.5 & 0.471 & 0.494 & 0.466 & 0.493 \\
        \multirow{2}{*}{\begin{tabular}[c]{@{}c@{}}Powerlaw slope modifying \\ the attenuation curve\end{tabular}} & Mean & -0.17 & -0.193 & -0.131 & -0.2 & -0.162 & -0.156 & -0.168 \\
        & SD & 0.145 & 0.144 & 0.149 & 0.141 & 0.15 & 0.15 & 0.149 \\
        \multirow{2}{*}{\begin{tabular}[c]{@{}c@{}}Goodness of fit \\ (reduced chi square)\end{tabular}} & Mean & 2.532 & 2.12 & 1.877 & 2.007 & 1.43 & 1.869 & 1.95 \\
        & SD & 1.544 & 1.267 & 1.485 & 1.46 & 1.338 & 1.059 & 1.408 \\ 
        \hline
    \end{tabular}
    }
    \label{tab:sed_parameters}
\end{table*}

\subsection{Star Formation Rate}

\begin{figure}
    \centering
    \includegraphics[width=1.0\linewidth]{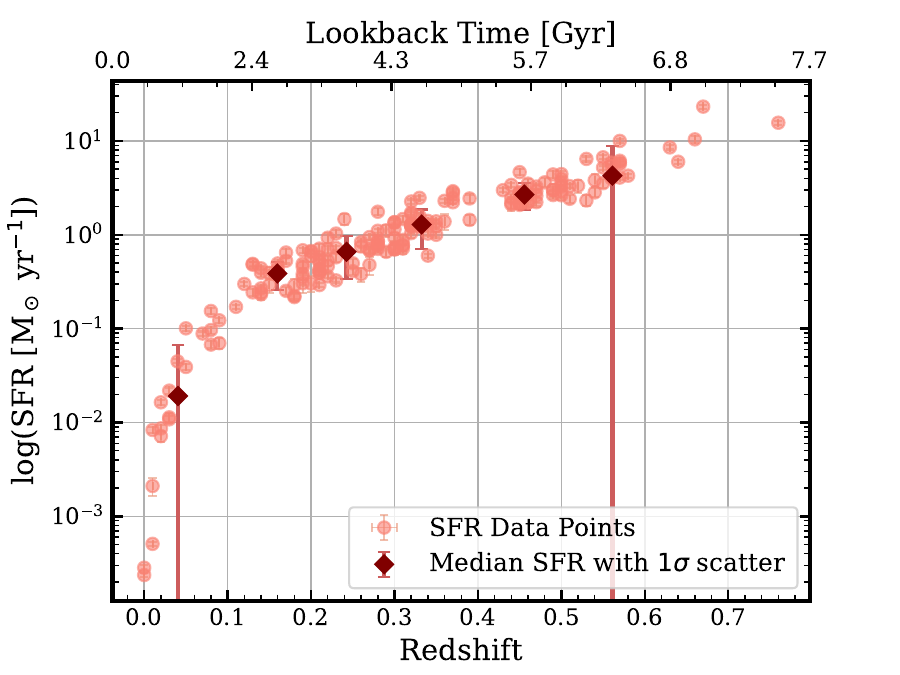}
    \caption{The SFR of the galaxies plotted against the redshift in red markers.}
    \label{fig:sfr_z_binned}
\end{figure}

Figure \ref{fig:sfr_z_binned} illustrates the evolution of the SFR as a function of redshift ($z$) for the observed galaxy sample, spanning from the local universe ($z \sim 0$) out to $z \approx 0.75$. The red circular markers represent the individual SFR measurements of the galaxies, complete with their associated uncertainties, while the black diamond markers denote the median SFR computed within distinct redshift bins. 

The plot exhibits a striking, monotonic increase in star formation activity with increasing redshift. In the lowest redshift regime ($z < 0.2$), the sample is heavily populated by galaxies exhibiting quiescent to low-level star formation, with the vast majority maintaining an SFR well below $1 \ M_{\odot} yr^{-1}$. This anchors the sample firmly within the established parameters of the local universe, consistent with the findings of \cite{brinchmann2004physical}. However, as the redshift increases, both the upper envelope of the individual data points and the binned median values rise substantially. By $z \sim 0.6$, the median SFR increases to approximately $4.5 \ M_{\odot} yr^{-1}$, while the individual scatter expands significantly.

This pronounced positive correlation between SFR and redshift is a well-documented physical reality of galaxy evolution. It is a direct reflection of the redshift evolution of the SFMS. As demonstrated by comprehensive multi-wavelength studies such as \cite{speagle2014highly} and \cite{whitaker2014constraining}, galaxies at higher redshifts inherently possessed larger cold gas reservoirs and higher dynamical gas fractions, driving much more vigorous star formation compared to their local counterparts of similar mass. The observed rise in our median values perfectly mirrors this secular evolution.

Furthermore, the overall trend observed in Figure \ref{fig:sfr_z_binned} traces the ascending slope of the cosmic star formation history. As reviewed by \cite{Madau_2014}, the volume-averaged star formation rate density of the universe increases steadily from the present day back to the "cosmic noon" at $z \sim 2$.

\subsection{Stellar Mass}

\begin{figure}
    \centering
    \includegraphics[width=1.0\linewidth]{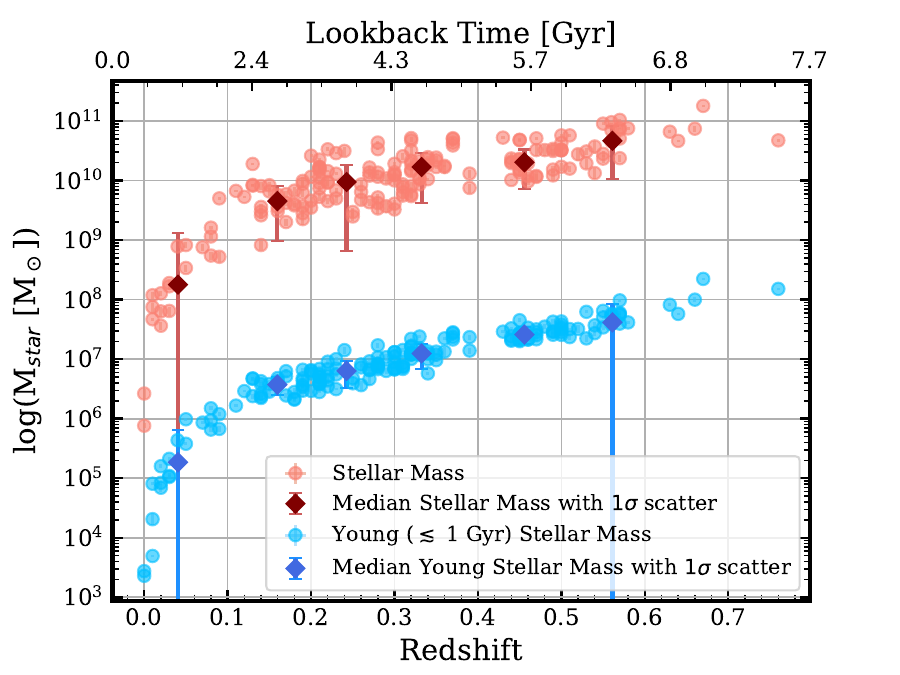}
    \caption{The logarithm of the stellar mass of the galaxies is plotted against the redshift in red markers and the logarithm of the young stellar mass of the galaxies is plotted against the redshift in blue markers.}
    \label{fig:mstar_z_binned}
\end{figure}

Figure \ref{fig:mstar_z_binned} presents the distribution of stellar mass and the mass of young stars in the galaxies as a function of redshift for the observed galaxy sample. The individual total stellar mass measurements are denoted by orange circular markers with their corresponding uncertainties, and young ($\lesssim$ 1 Gyr) stellar mass is denoted by blue circular markers, while the red diamond markers signify the median stellar mass and the blue diamond signify the median young stellar mass calculated within distinct redshift intervals with the bars showing $1 \sigma$ scatter. The y-axis traces the stellar mass on a logarithmic scale.

The plot displays a positive correlation between the detected stellar mass and redshift. In the local universe ($z < 0.1$), the survey is sensitive to a broad dynamic range of galaxy masses, successfully detecting low-mass dwarf galaxies with stellar masses down to $M_* \sim 10^6 - 10^8 \ M_{\odot}$ with mass of young stars down to $M_* \sim 10^3 \ M_{\odot}$. However, as the redshift increases, the lower boundary of the detectable mass distribution rises sharply. Consequently, the median stellar mass of the sample shifts dramatically, ascending from approximately $10^{8.2} \ M_{\odot}$ at $z \sim 0.05$ to roughly $10^{10.7} \ M_{\odot}$ by $z \sim 0.55$, where the sample becomes almost entirely populated by massive galaxies.

It is a classic manifestation of Malmquist bias within a flux-limited survey. Because the apparent brightness of a galaxy drops with the square of its luminosity distance, low-mass (and therefore intrinsically fainter) galaxies rapidly fall below the signal-to-noise detection thresholds of the observational catalogues at higher redshifts. As extensively detailed in foundational mass-completeness studies, such as \cite{pozzetti2010zcosmos} and \cite{ilbert2013mass}, the sharply rising lower envelope in this parameter space physically delineates the evolving mass completeness limit of the survey. 

At higher redshifts ($z > 0.4$), the survey only detects the most massive and luminous galaxies, typically those residing near or above the characteristic Schechter turnover mass. This observational constraint is highly consistent with deep-field Galaxy Stellar Mass Function (GSMF) measurements by \cite{davidzon2017cosmos2015}. Their work demonstrates that while low-mass galaxies are numerically dominant at all cosmic epochs, identifying systems below $10^{10} \ M_{\odot}$ at $z > 0.5$ requires extreme observational depths. Therefore, Figure \ref{fig:mstar_z_binned} effectively maps the observational selection function of the current sample, emphasising that any comparative evolutionary analysis at $z > 0.4$ must be restricted to the high-mass ($>10^{10} \ M_{\odot}$) regime to ensure statistical mass completeness.

\begin{figure}
    \centering
    \includegraphics[width=1.0\linewidth]{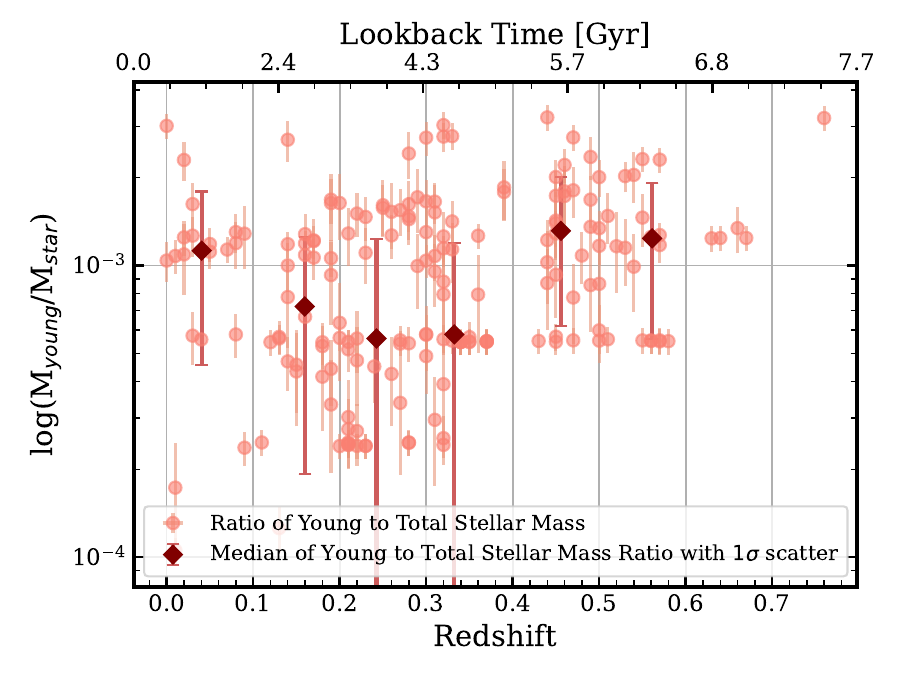}
    \caption{The logarithm of the ratio of the mass of the young stars to the total mass of the galaxies is plotted against the redshift in red markers.}
    \label{fig:ratio_young_total_z}
\end{figure}

Figure \ref{fig:ratio_young_total_z} presents the evolution of the logarithmic ratio between the young stellar mass and the total stellar mass as a function of redshift. Individual galaxy measurements are marked by red circles with corresponding error bars, while the black diamonds represent the median mass fraction calculated within discrete redshift bins. 

Unlike the strong positive correlations observed in the absolute SFR and absolute mass distributions, this relative mass fraction exhibits a remarkably flat evolutionary trend. The binned median values fluctuate only marginally, remaining tightly anchored near $10^{-3}$ (or $0.1\%$) across the entire surveyed redshift range ($0 < z < 0.6$). While the individual data points display intrinsic scatter, ranging from $10^{-4}$ to slightly above $10^{-3}$, the overall sample consistently maintains a very low ratio of newly formed stars to total assembled mass.

This constant fractional mass assembly demonstrates that while the absolute scales of star formation and total mass are subject to strong observational selection limits at higher distances, the intrinsic mode of stellar mass growth remains strictly uniform across the detected sample. A young mass fraction of $\sim 0.1\%$ is a classic, quantitative signature of mature, secularly evolving galaxies situated firmly on the SFMS. As established by \cite{noeske2007star} and further mapped by \cite{elbaz2011goods}, main-sequence galaxies grow through continuous, self-regulated gas accretion over cosmological timescales, rather than through violent, chaotic, merger-driven starbursts. If this UV-selected sample were heavily contaminated by starbursting systems, the young stellar mass fraction would transiently spike to significantly higher values (e.g., $1\%$ to $10\%$). Instead, our data confirms that recent star formation is merely providing a continuous, light addition to a predominantly old, pre-established stellar backbone.

Furthermore, this empirical result robustly validates the parametric assumptions embedded within our CIGALE SED modelling. The observed mass ratio perfectly mirrors the constrained late-burst mass fractions (set at $0.00$ to $0.01$) utilised within the \texttt{sfhdelayed} module. It confirms that the observed FUV flux, despite being exclusively sensitive to the youngest, most massive stars \cite{kennicutt2012star}, is accurately and physically reproduced by models where the recent star formation episode is a highly localised, minor contributor to the global baryonic mass budget. Ultimately, this invariant ratio underscores the steady-state nature of mass assembly for intermediate-to-high mass galaxies over the past several billion years of cosmic history.

\section{Conclusion} \label{sec:conc}
We present the FUV catalogue of the ELAIS N1 deep field from UVIT aboard the AstroSat. We give 2 catalogues at $3 \sigma$ and $5 \sigma$ detection limits. The $3 \sigma$ and $5 \sigma$ detections are $25.69$ and $25.13$ respectively. 

In this study, we presented an extensive photometric analysis of the ELAIS N1 deep field utilising high-resolution FUV observations from the AstroSat/UVIT instrument. By reaching a $3\sigma$ detection limit of $25.69 \ m_{AB}$ in the F154W filter, we constructed a robust, deep FUV source catalogue. To comprehensively understand the evolutionary state of these galaxies, we cross-matched our UV detections with a wealth of multi-wavelength ancillary data, encompassing optical photometry and spectroscopy from SDSS/BOSS, mid-infrared data from the SWIRE survey, and highly accurate photometric redshifts derived from the LoTSS deep field survey. After isolating AGN using established multi-wavelength criteria, our final sample consists of star-forming galaxy population within this field.

To derive the fundamental physical properties of these galaxies, we performed rigorous SED modelling using CIGALE. By employing a delayed star formation history with an optional late burst, alongside sophisticated treatments for dust attenuation \citep{calzetti2000dust} and emission \citep{Dale2014}, we successfully constrained the global SFR, total stellar masses, and young stellar mass fractions for our sample. 

The SFR of the sample increases monotonically with redshift, rising from median values well below $1\,\mathrm{M_{\odot}\,yr^{-1}}$ at $z < 0.2$, consistent with local universe measurements of \citet{brinchmann2004physical}, to approximately $4.5\,\mathrm{M_{\odot}\,yr^{-1}}$ at $z \sim 0.6$. This trend is in    excellent agreement with the well-established redshift evolution of the SFMS \citep{speagle2014highly, schreiber2015herschel} and directly traces the ascending slope of the cosmic SFR density \citep{Madau_2014}.

The detected stellar mass range  shifts from low-mass systems ($M_* \sim 10^{6}$--$10^{8}\,\mathrm{M_{\odot}}$) in the local universe to exclusively massive galaxies ($M_* \gtrsim 10^{10}\,\mathrm{M_{\odot}}$) at $z > 0.4$, a direct manifestation of Malmquist bias in a flux-limited UV survey. This behaviour is consistent with mass-completeness analyses of deep surveys such as zCOSMOS \citep{pozzetti2010zcosmos} and COSMOS2015 \citep{ilbert2013mass, davidzon2017cosmos2015}. Any evolutionary comparison at $z > 0.4$ within this sample must therefore be restricted to the high-mass regime, where the Schechter function turnover mass \citep[$M^* \sim 10^{10.5}$--$10^{10.8}\,\mathrm{M_{\odot}}$;][]{baldry2012galaxy} anchors the galaxy population.

Most notably, the ratio of young stellar mass to total stellar mass remains remarkably flat at $\sim 10^{-3}$ ($\sim 0.1$\,per\,cent) across the entire surveyed redshift range $0 < z < 0.6$, despite the strong trends observed in the absolute SFR and stellar mass distributions. This constancy is a quantitative signature of secularly evolving, SFMS galaxies growing through continuous, self-regulated gas accretion \citep{noeske2007star, elbaz2011goods}, rather than through starburst-driven episodic assembly \citep{daddi2010different, rodighiero2011lesser}. The observed mass fraction directly validates the constrained late-burst mass fractions (set to $0.00$--$0.01$) within the \texttt{sfhdelayed} CIGALE module. This result is consistent with the picture of \citet{Madau_2014}, wherein the bulk of cosmic stellar mass was assembled at the peak of cosmic star formation ($z \sim 2$), with present-day galaxies adding only marginal new stellar mass relative to their total assembled budget.

These results demonstrate that deep AstroSat/UVIT FUV observations, combined with the rich multiwavelength ancillary data of the ELAIS N1 field and self-consistent SED modelling with CIGALE, provide a physically consistent and observationally robust picture of galaxy evolution across $0 < z \lesssim 0.76$. Future work extending this analysis to higher redshifts, studying the UV luminosity function, and comparing with the radio-detected SFG populations catalogue by uGMRT \citep{Arnab2019_2, sinha2022deep} and LOFAR \citep{Sabater_2021} in this field will further constrain the relationship between UV-traced and bolometric star formation activity across cosmic time.

\section*{Acknowledgments}


PC is funded by ISRO under the Sponsored Research Project no. DS 2B-13013(2)4/2021/Sec-2. This publication uses the data from the AstroSat mission of the Indian Space Research Organisation (ISRO), archived at the Indian Space Science Data Centre (ISSDC). This work uses observations made with the Spitzer Space Telescope, which was operated by the Jet Propulsion Laboratory, California Institute of Technology under a contract with NASA. Support for this work was provided by NASA through an award issued by JPL/Caltech. Funding for the SDSS and SDSS-II has been provided by the Alfred P. Sloan Foundation, the Participating Institutions, the National Science Foundation, the U.S. Department of Energy, the National Aeronautics and Space Administration, the Japanese Monbukagakusho, the Max Planck Society, and the Higher Education Funding Council for England. The SDSS Web Site is \url{http://www.sdss.org/}. The SDSS is managed by the Astrophysical Research Consortium for the Participating Institutions. The Participating Institutions are the American Museum of Natural History, Astrophysical Institute Potsdam, University of Basel, University of Cambridge, Case Western Reserve University, University of Chicago, Drexel University, Fermilab, the Institute for Advanced Study, the Japan Participation Group, Johns Hopkins University, the Joint Institute for Nuclear Astrophysics, the Kavli Institute for Particle Astrophysics and Cosmology, the Korean Scientist Group, the Chinese Academy of Sciences (LAMOST), Los Alamos National Laboratory, the Max-Planck-Institute for Astronomy (MPIA), the Max-Planck-Institute for Astrophysics (MPA), New Mexico State University, Ohio State University, University of Pittsburgh, University of Portsmouth, Princeton University, the United States Naval Observatory, and the University of Washington. This work has made use of data from the European Space Agency (ESA) mission {\it Gaia} (\url{https://www.cosmos.esa.int/gaia}), processed by the {\it Gaia} Data Processing and Analysis Consortium (DPAC, \url{https://www.cosmos.esa.int/web/gaia/dpac/consortium}). Funding for the DPAC has been provided by national institutions, in particular the institutions participating in the {\it Gaia} Multilateral Agreement. This research made use of Photutils, an Astropy package for detection and photometry of astronomical sources \citep{larry_bradley_2024_13989456}.

\section*{Data Availability}

The ELAIS N1 FUV 3$\sigma$ \& $5\sigma$ source catalogues and results from SED fitting with CIGALE will be available as online tables.



\bibliographystyle{mnras}
\bibliography{references} 






\bsp	
\label{lastpage}
\end{document}